\documentclass[10pt,letterpaper,twocolumn]{article} %% two column, final layout

\usepackage{ol2}
\usepackage[draft,implicit=false]{hyperref}
\usepackage{amsmath,amssymb,graphicx}
\graphicspath{ {Figures/} }
\begin{document}

\twocolumn[ %% activate for two-column option

\title{Phase sensitive amplification in silicon photonic crystal waveguides}

%% For REVTeX it is possible to automate superscript and e-mail callouts with the superscriptaddress option; see REVTeX4 documentation.

\author{Yanbing Zhang,$^{1,*}$Chad Husko,$^1$Jochen Schr\"{o}der,$^1$Simon Lefrancois,$^1$ \\Isabella H. Rey,$^2$ Thomas F. Krauss,$^{2,3}$ and Benjamin J. Eggleton$^1$}

\address{
$^1$Centre for Ultrahigh bandwidth Devices for Optical Systems (CUDOS), Institute of Photonics and Optical Science (IPOS), \\ School of Physics, University of Sydney, NSW 2006, Australia
\\
$^2$SUPA, School of Physics and Astronomy, University of St. Andrews, Fife, KY169SS, UK \\
$^3$School of Physics, University of York, York, YO10 5DD, UK \\
$^*$Corresponding author: y.zhang@physics.usyd.edu.au
}

\begin{abstract}We experimentally demonstrate phase sensitive amplification (PSA) in a silicon photonic crystal waveguide based on pump-degenerate four-wave mixing. An 11 dB phase extinction ratio is obtained in a record compact 196 $\mu m$ nanophotonic device due to broadband slow-light, in spite of the presence of two-photon absorption and free-carriers. Numerical calculations show good agreement with the experimental results.\end{abstract}


 ] %% activate for two-column option

\noindent Phase sensitive amplification (PSA) is an extremely attractive function for integrated all-optical classical and quantum signal processing due to a wide range of potential applications. In classical communications, PSA is a core technology for noiseless amplification and optical phase regeneration \cite{PSAinfiber2004,slavik2010all,umeki2013line,tong2011towards}, while in quantum processing it is useful for generating squeezed states of light \cite{levandovsky1999,caves1982quantum} or quantum state translation via Bragg scattering \cite{mcguinness2010,clark2013QST}. Despite this progress, the majority of experiments have been in fiber \cite{slavik2010all,tong2011towards} with only a handful of investigations of PSA in centimetre long $\chi^{(2)}$ \cite{lee2009phase,puttnam2011phase,umeki2013line} or $\chi^{(3)}$ devices \cite{kang2013experimental,neo2013phase}.

Simultaneous to the these developments in large-scale all-optical signal processing, the miniaturization and integration of optical devices onto monolithic photonic chips is rapidly progressing \cite{leuthold2010,politi2009integrated}. In order to meet these demanding requirements, the nonlinear elements in current bulk systems must be significantly scaled down in size. To this end, the strong nonlinear response of materials with high refractive index such as semiconductors \cite{leuthold2010} and chalcogenide glasses \cite{eggleton2011chalcogenide} enable tight confinement with waveguides centimetres, or even millimetre, scales in lieu of meters of fiber. Silicon is an especially compelling material due to compatibility with CMOS electronics and consequent scope for hybrid opto-electronic chips with unprecedented information processing capabilities \cite{leuthold2010}. A drawback of the silicon platform in the near-infrared 1.55~$\mu$m band is the presence of two-photon absorption (TPA) and accompanying free-carriers \cite{yin2007impact,foster2006broad}, which limit the desirable $\chi^{(3)}$ Kerr nonlinearity underlying many applications in ultrafast all-optical signal processing.

In this Letter, we demonstrate PSA in a record short 196~$\mu$m silicon device. By taking advantage of the slow-light enhanced nonlinearity and dispersion engineering capabilities of photonic crystal waveguides (PhCWG) \cite{li2008systematic,monat2010slow}, we achieve a phase extinction ratio (PER) of 11~dB in a pump-degenerate four-wave mixing (FWM) scheme. While TPA and free-carrier effects in the silicon material restrict the maximum phase sensitive gain (PSG) \cite{foster2006broad}, we demonstrate that the phase extinction ratio \cite{slavik2010all}, the crucial parameter for phase regeneration, remains essentially unaffected. 

% ================================================================================================
\begin{figure}[htb]
\begin{center}
%includegraphics[width=8.4cm]{Fig1_FWM_setup}
\includegraphics[width=8cm]{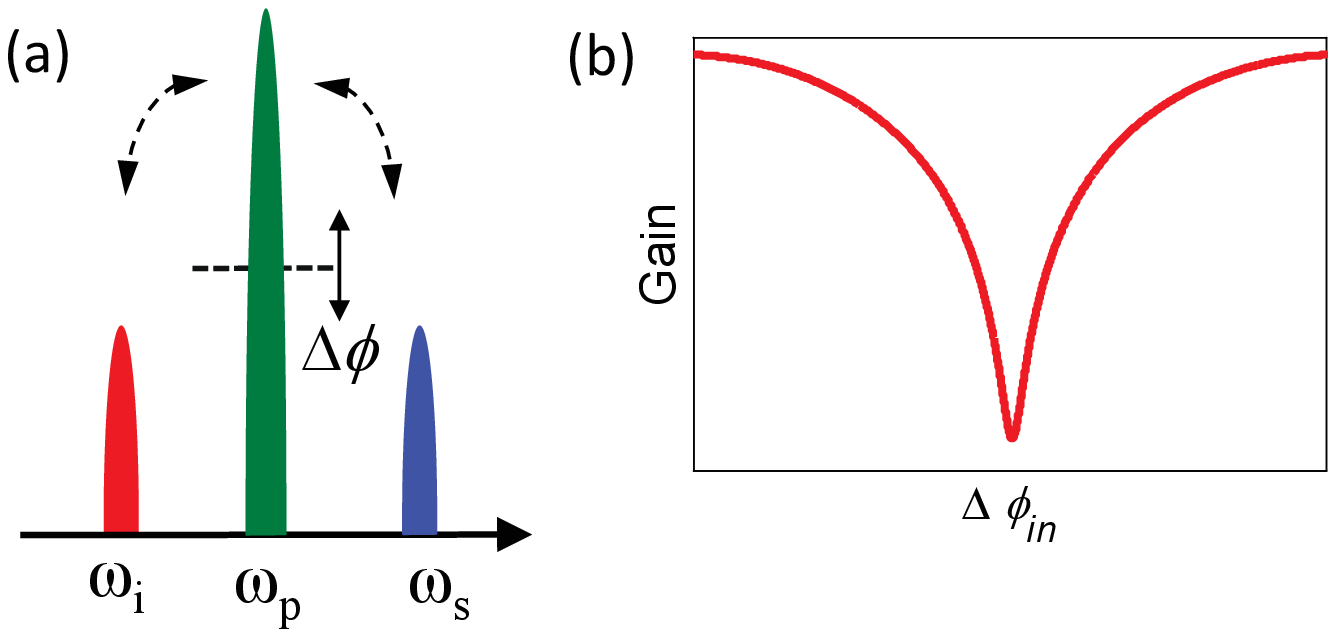}
\includegraphics[width=8.4cm]{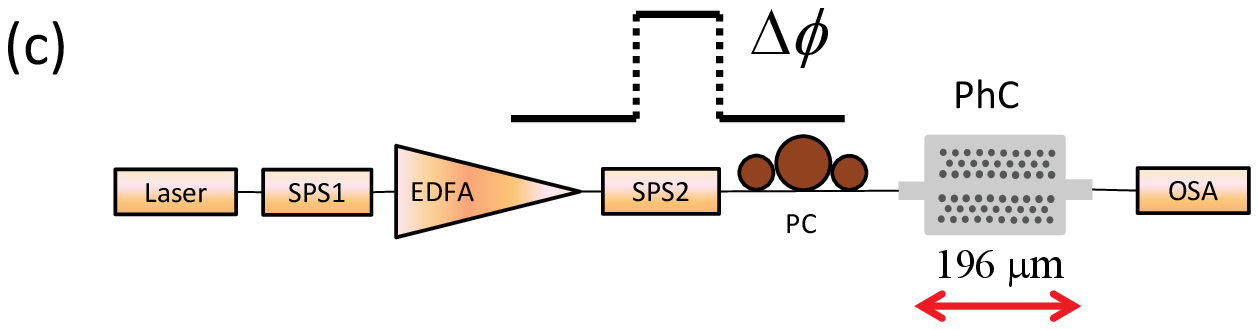}
\caption{(a) Schematic of the pump-degenerate PSA based on FWM. (b) Output intensity of the signal vs input phase detuning. (c) PSA experimental setup. Symbols are defined in the text.}
\label{fig:setup}
\end{center}
\end{figure}
% ================================================================================================

Two common architectures for phase-sensitive amplification using $\chi^{(3)}$ materials are signal-degenerate \cite{slavik2010all} and pump-degenerate \cite{tong2011towards} FWM. Fig.~\ref{fig:setup}(a) depicts the principle of phase-sensitive pump-degenerate FWM we employ in our experiments. In this configuration a pump (green) at frequency $\omega_{\rm p}$ is converted to an idler $\omega_{\rm i}$ (red) and a signal $\omega_{\rm s}$ (blue) given by the energy conservation relation $\omega_{\rm s}+\omega_{\rm i} = 2\omega_{\rm p}$. When the pump, signal and idler waves are seeded simultaneously at the input port, the FWM process is phase sensitive with the energy transfer between three waves depending on their relative phase. Therefore, we can control the signal amplification by detuning the phases of the waves, e.g. the pump phase $\Delta\phi$ in Fig. \ref{fig:setup}(b).

% ================================================================================================
\begin{figure}[*htb]
\centering
%\begin{tabular}{cc}
%\begin{minipage}{120pt}
%\centerline{\includegraphics[width=130pt]{transmission}}
%\end{minipage}
%\begin{minipage}{120pt}
%\centerline{\includegraphics[width=130pt]{power_in_out}}
%\end{minipage}
%\end{tabular}
\includegraphics[width=7cm]{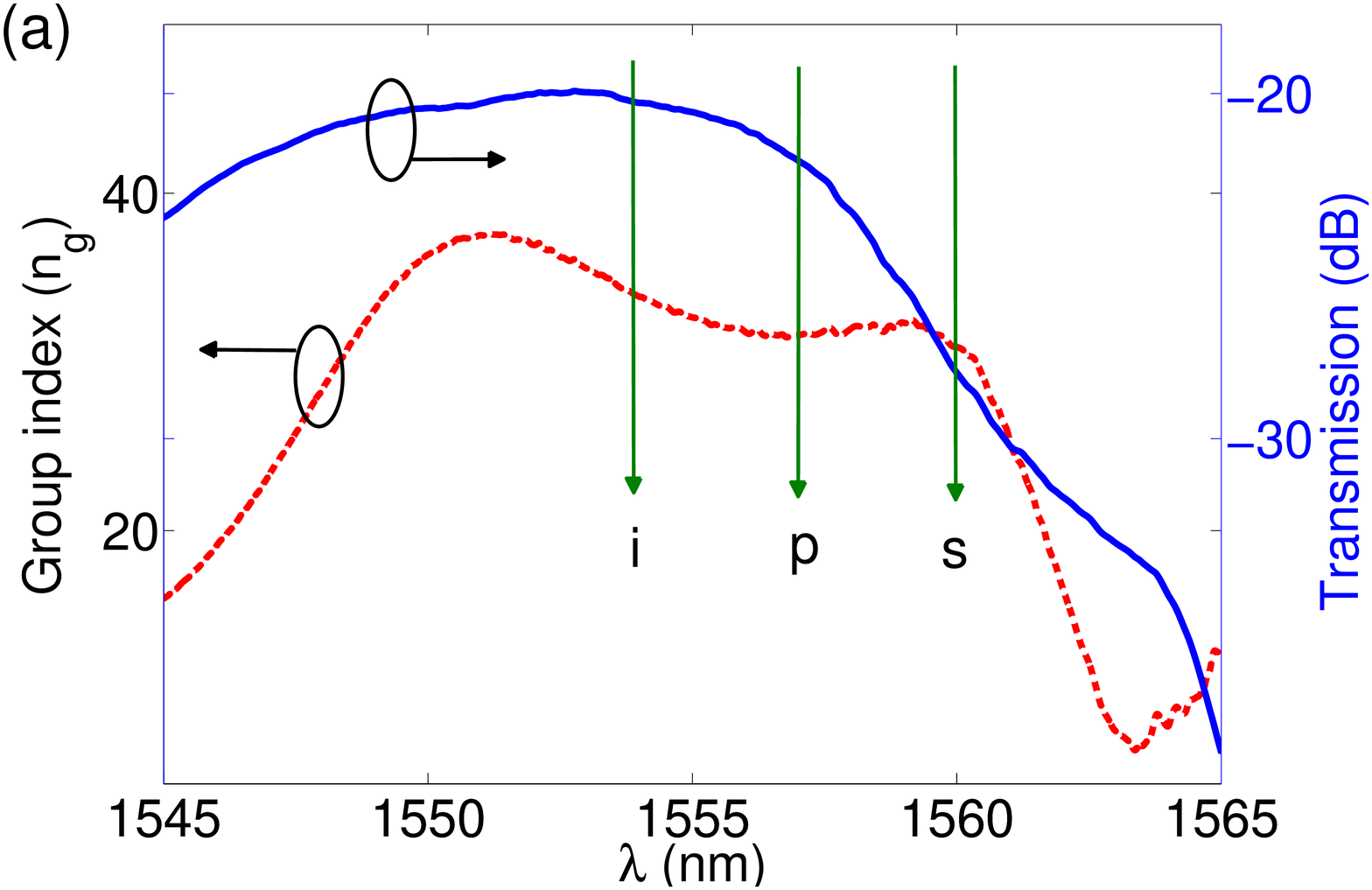}
\includegraphics[width=7cm]{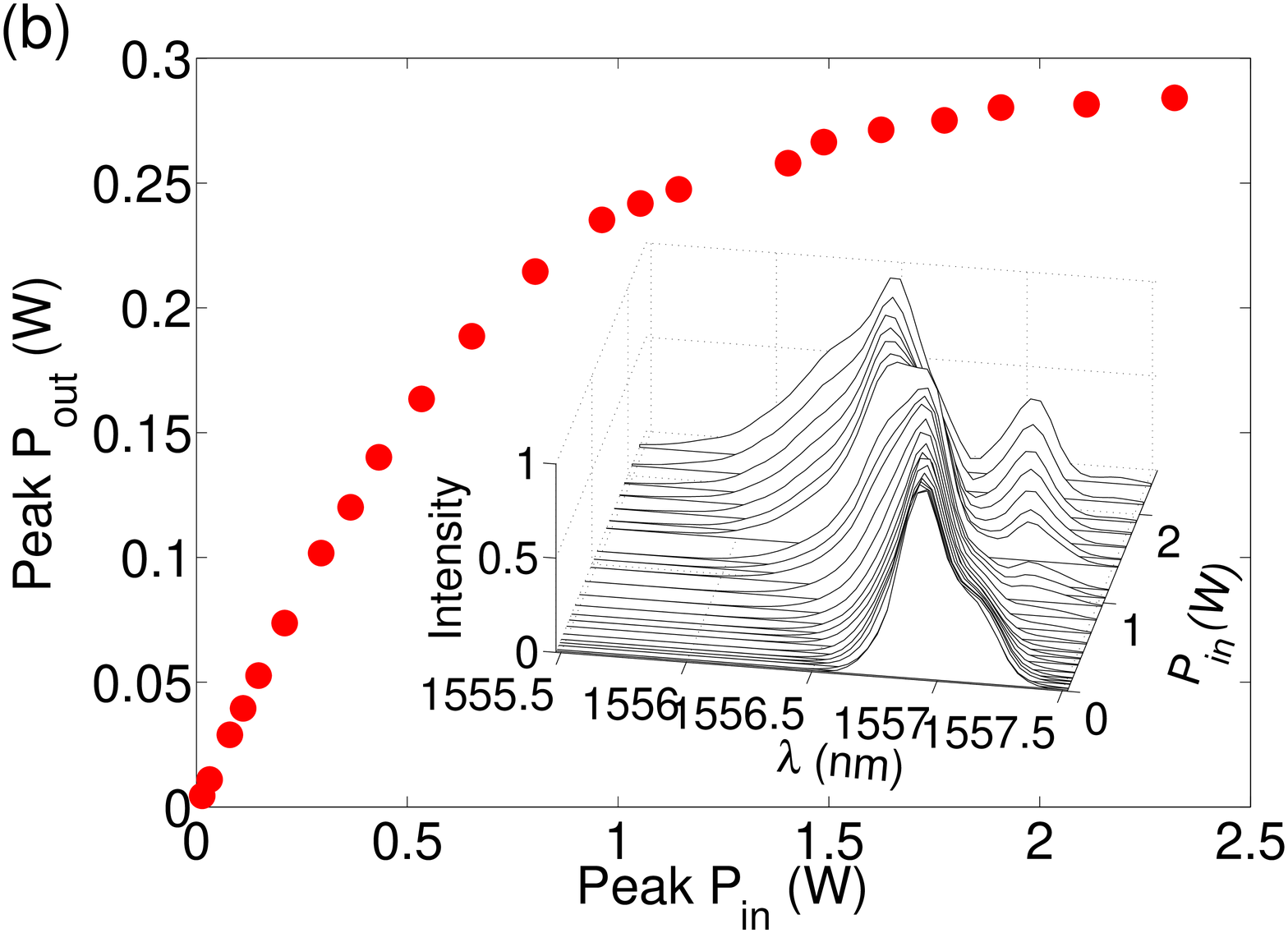}
\caption{(a) Measured linear transmission spectrum (fiber to fiber) and group index of the PhC waveguide, where arrows indicate the position of the idler, signal and pump. (b) Measured P$_{\rm in}$ vs. P$_{\rm out}$ of the pump showing saturation due to TPA and FCA. Inset: normalized output spectra of the pump only as a function of the input peak power.}
\label{fig:properties}
\end{figure} 
% ================================================================================================

Fig. \ref{fig:setup}(c) shows the PSA experimental setup. Since the three waves need to be synchronized and phase locked, we generate the signal/pump/idler waves by spectrally slicing the broadband (30~nm) output of a mode-locked Erbium-doped fiber laser (38.6~MHz) using a spectral pulse shaper (SPS1 - Finisar Waveshaper). The resulting idler/pump/signal waves are Gaussian shaped with centre wavelengths at 1554/1557/1560~nm and time durations 8/15.2/8~ps. To avoid FWM inside the EDFA, the signal and idler were delayed by 15/-15~ps with respect to the pump pulse. A second spectral processor (SPS2) is then used to bring the pulses back together and to adjust the peak power and phase of the three waves. The polarization of the light is aligned to the TE slow-light mode of the waveguide by a polarization controller (PC). The input and output powers to the chip were monitored by a power meter (PM) and the output spectra were recorded using an optical spectrum analyzer (OSA).

Figure~\ref{fig:properties}(a) shows the linear transmission (including coupling losses) and group index of the TE mode of the silicon PhCWG. The geometric design of the PhCWG is given in Ref. \cite{li2011FWM}. The dispersion engineered silicon PhC waveguide has a group index of $ n_g \approx 32$ over a~10 nm bandwidth with negligible dispersion for the pulses considered here (dispersion length $L_{\rm \beta}\sim 14$ cm). Assuming the coupling loss is 9/10dB at the input/output port, the linear propagation loss and the effective nonlinear coefficient were inferred to be around 150~dB/cm at 1557~nm and $\gamma_{\rm eff}=\gamma_{\rm PhC}S^2$=4300(W-m)$^{-1}$ with the slow-down factor $S=n_g/n_{si} = 32/3.45$ \cite{li2011FWM}. Figure \ref{fig:properties}(a) also indicates the position of the three waves with pump at 1557~nm exhibiting anomalous dispersion $\beta_2=-1.7~\mathrm{ps}^2/\mathrm{mm}$. Note the signal (1560~nm) has a slightly larger propagation loss.

Before the PSA experiments, we first investigated the role of TPA/FCA on the pump propagation. Figure \ref{fig:properties}(b) shows the measured on-chip output peak power versus coupled input peak power. At large input power, the output power saturates due to strong TPA with the optically generated free-carriers inducing absorption (FCA) further limiting the peak power. The inset shows the corresponding normalized output spectra of the pump as a function of the input peak power. While the TPA attenuates the pulses symmetrically, the time delayed free-carriers affect the trailing edge of the pulse, leading to the asymmetric blue-shift in the pump spectra due to free-carrier dispersion (FCD) \cite {monat2009,husko2013soliton}. With this knowledge we set the on-chip pump peak power in our first experiment to $P_p =1.5$~W to avoid large losses to TPA and FCA. We inject the signal and idler waves with much smaller peak power levels of 40~mW and 10~mW, respectively, well below the nonlinear threshold. 

\begin{figure}[htb]
\centering
\includegraphics[width=8cm]{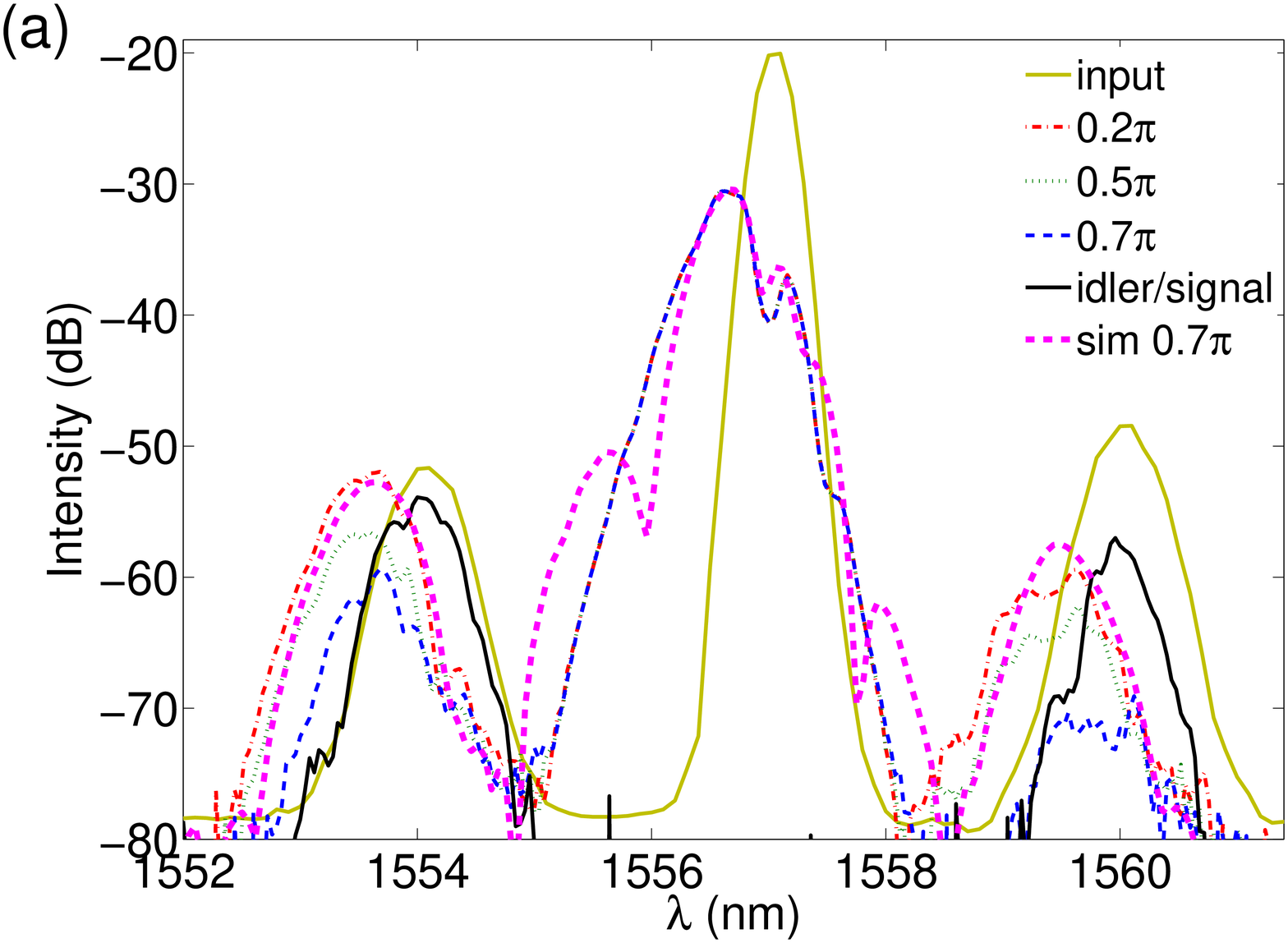}
\includegraphics[width=8cm]{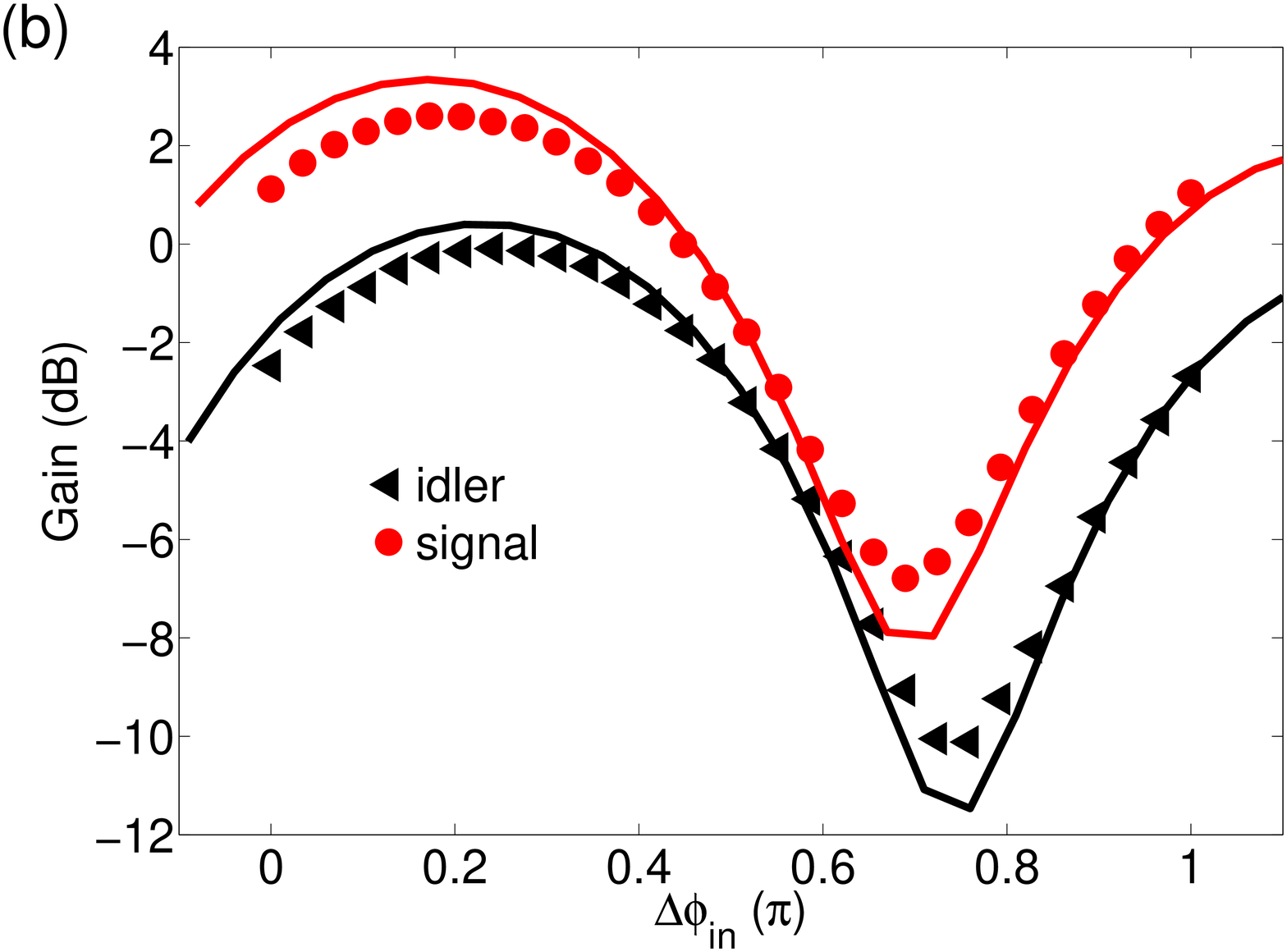}
\caption{(a) Input spectrum (solid line), output spectra with relative phases (dashed lines), signal only (light colored) and the simulated output spectrum of the PSA process (bold dash line). (b) Experimentally observed signal (triangle) and idler gain (dot), and numerically (solid) versus phase detuning ($\Delta\phi$) at peak power of 1.5 W.}
\label{fig:gain}
\end{figure}

In order to characterize the PSA, we measured the on-chip signal/idler gain by varying the pump phase ($\Delta\phi$) while keeping the phases of the signal/idler waves constant at SPS2. Figure \ref{fig:gain}(a) shows an overlay of output spectra measured at different pump phase shifts of $0.2\pi$, $0.5\pi$, and $0.7\pi$, as well as the unamplified signal/idler waves with no pump. We observe a clear variation in the signal and idler intensities for different values of the pump phase indicating the PSA process is performing as expected. We note the signal intensity is lower than the idler due to the higher propagation loss at the signal wavelength (see Fig. \ref{fig:properties}(a)) even though we inject a higher power.

% ================================================================================================
% ================================================================================================
The spectra of Fig. \ref{fig:gain}(a) also highlight several features of PSA in silicon. In contrast to earlier work in glass materials \cite{PSAinfiber2004,neo2013phase}, our demonstration in the semiconductor silicon material includes TPA and free-carriers modulating the PSA process. In particular, free carriers generated by the strong pump, blue-shift the spectra of all three waves (dotted lines), readily seen when comparing the output to the input spectra. While the blue-shift in semiconductors is known from other nonlinear semiconductor experiments \cite{monat2009,li2011FWM,husko2013soliton}, this is the first investigation into the role of TPA and free-carriers on the PSA process.

Fig.~\ref{fig:gain}(b) summarizes the experimental signal and idler gain as a function of relative phase detuning $\Delta\phi$ over one period. Note that the period of the PSA curve is $\pi$, due to the factor of two in the FWM phase matching relation for the phase of the pump. We define the on-chip gain as: $G = \frac{ \int{ P_s(out)(\lambda) } }{ \int{P_s(out-no pump)(\lambda) }}$, the ratio of the amplified signal power with the pump wave switched on over the unamplified signal power without the pump at the output of the waveguide. We employ this definition for simple comparison at the output OSA. We integrate over the signal and idler spectral width to account for self and cross-phase modulation broadening shown in Fig.~\ref{fig:gain}(a). Experimentally, a maximum signal (idler) gain of 0~(2.7)~dB is obtained at 0.68~(0.75)$\pi$ and a minimum gain of -10~(-6.8)~dB occurs at 0.18~(0.25)$\pi$, yielding an extinction ratio of 10~dB (9.5~dB) for the signal/idler at this power level. We believe the maximum and minimum gain curves are asymmetric due to TPA/FCA limiting the maximum gain, and attribute the difference in gain to the different transmission properties at these wavelengths. Similarly the small phase offset between the signal and idler gain can be explained by free-carriers inducing slightly different FCD (blue shift).

In Fig.~3, we confirm these measurements by numerical solution of the nonlinear Schr\"{o}dinger equation
(NLSE) with split-step Fourier method (SSFM) including up to the 4th order dispersion, TPA and free
carriers \cite{yin2007impact}. The parameters used in the simulation of the PSA experiment are given in
Table~\ref{tab:params}. Note that our model uses slightly different values for the TPA coefficient 
($1.2\times\alpha_{\mathrm{TPA}}$) and free carrier dispersion ($2.5\times k_c$) 
compared to references to most accurately
model our experimental results. Here, the linear propagation loss is derived directly from the measured
transmission spectrum in Fig. 2(a). The used pump pulse is measured from frequency resolved optical gating
(FROG) and the signal/idler are assumed as Gaussian pulses. A free carrier lifetime ($\tau_c$) is set as
300~ps \cite{monat2009}. The estimated nonlinear effect in the 60 $\mu m$-length input and output
nanowire is also negligible compared to the PhC waveguide. The simulation results show good agreement with the experimental observation. Note that the absolute phase in experiment and simulation is ambiguous. Therefore the phase origin of the simulated gain curves were shifted to match the experiments. 

\begin{table}[h]
{\bf\caption{Simulation parameters used in Fig. 3 \cite{monat2009,yin2007impact}}}\label{tab:params}\begin{center}
\begin{tabular}{lp{1.2in}lp{1.2in}}\hline
$P_p$&1.5 W &$P_s/P_i$ &42/10.4 mW \\
$t_p$&FWHM=15.2 ps&$t_s/t_i$ &FWHM=5/6 ps\\
$\beta_2$&$-1.7\times10^{-21}\mathrm{s}^2\mathrm{/m}$&$\beta_3$&$-3\times10^{-33}\mathrm{s}^3\mathrm{/m}$\\
$\beta_4$&$3\times10^{-46}$ s$^4$/m & {\sl L$_{\rm PhC}$}&196 $\mu$m \\
$n_{2}$ &$6\times10^{-18}\mathrm{m}^2\mathrm{/W}$&$\alpha_{\mathrm{TPA}}$&$1.2\times10^{-11}{\mathrm{m/W}} \mu m$\\
$\sigma_{\mathrm{FC}}$ &$1.45\times10^{-21}$m$^2$ & $k_{\mathrm{c}}$&$3.4\times10^{-27}$m$^3$\\
$A_{\mathrm{eff}}$&$0.5 \mu \mathrm{m}^{2}$ & $\tau_{\mathrm{c}}$&300~ps\\
\hline
\end{tabular}
\end{center}
\end{table}

\begin{figure}[*htb]
\centering
\includegraphics[width=7cm]{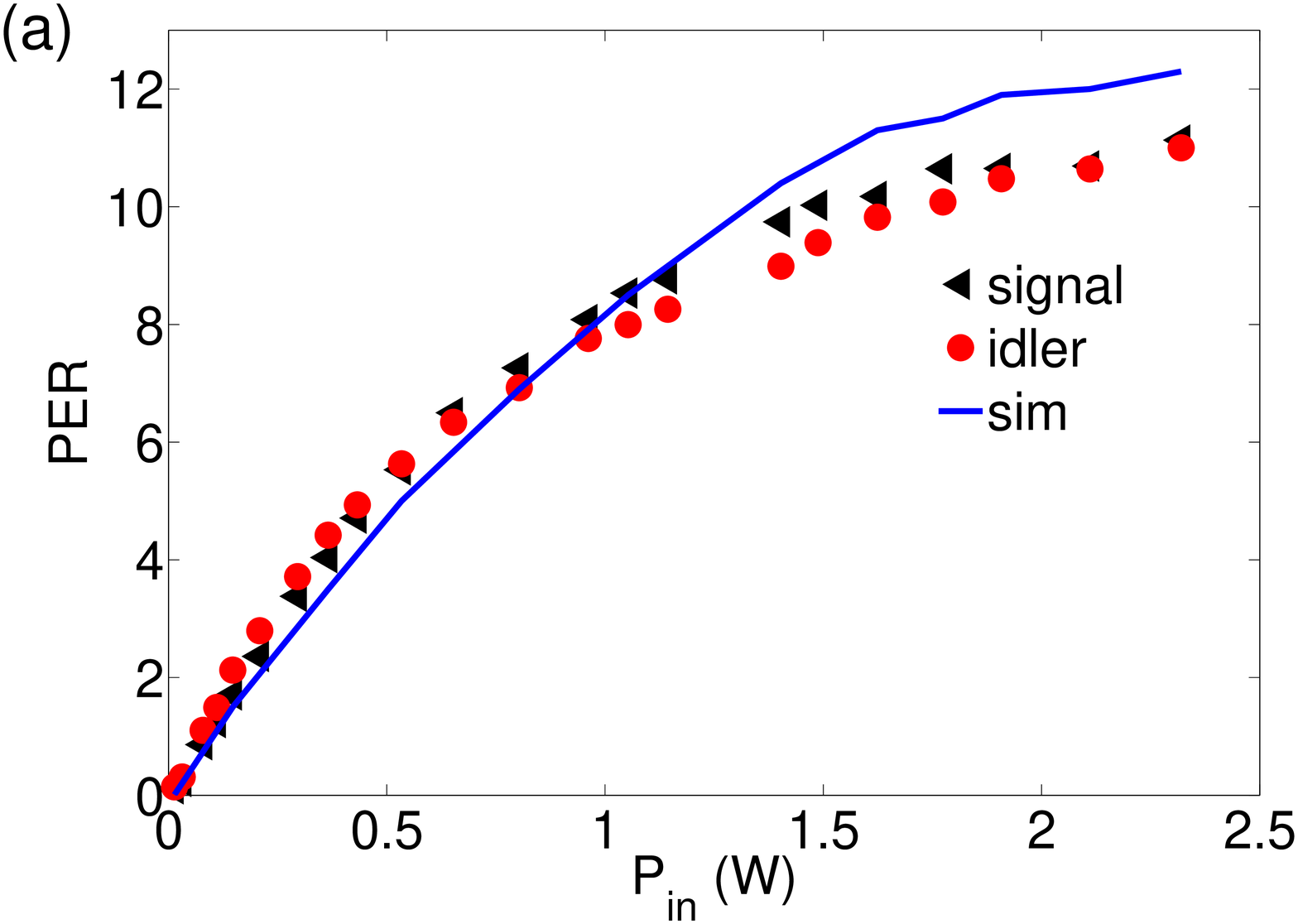}
\includegraphics[width=7cm]{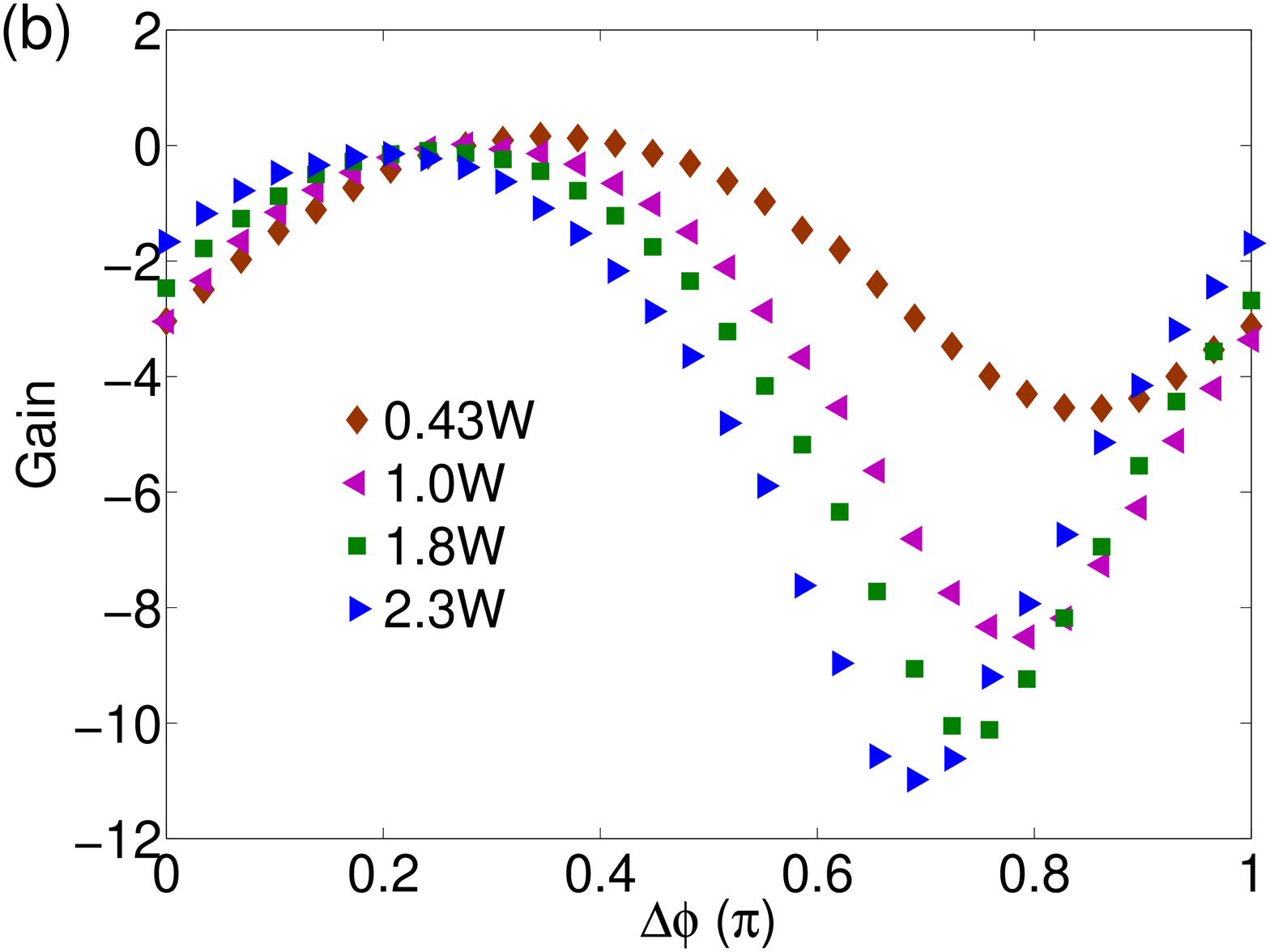}
\caption{(a) The phase extinction ratios (PER) of the signal (triangle) and idler (dot), and simulated (solid line) gain as a function of the pump peak power. (b) PSA phase dependent gain curves at varying pump powers.}
\label{fig:gain_power}
\end{figure}

To further explore the dynamics of the PSA process in the presence of TPA and free carriers, we repeated the experiment in Fig.~\ref{fig:gain} for various input pump powers. Figure~\ref{fig:gain_power}(a) presents the results of these measurements and unveils additional aspects unique to PSA in silicon in addition to the aforementioned blue-shifted spectra. First, we achieved a maximum extinction ratio of 11~dB (11.5 dB) for the signal (idler) at a pump power of 2.3~W. Importantly the extinction ratios of both the signal and idler are consistent throughout the entire range of power detuning as shown in Fig.~\ref{fig:gain_power}(a). Based on previous demonstration, this PER is sufficient for phase regeneration \cite{slavik2010all}. Second, the increase of the extinction ratio slows down at 1.0~W due to the saturating nature of the nonlinearity as indicated earlier in the $P_{in}-P_{out}$ curve of Fig. \ref{fig:properties}(b).

Figure~\ref{fig:gain_power}(b) shows the phase sensitive gain of the signal wave at four different pump powers from Fig.~\ref{fig:gain_power}(a). One observes a clear shift of the phase position of the maxima/minima as a function of input power. The underlying physical reasoning is a larger pump power causing a larger nonlinear phase shift ($\phi_{NL}$) adjusting the phase matching condition. The maximum gain occurs at $\pi/2-\phi_r$, where $\phi_r=\left(\phi_{NL}-\phi_{\beta}\right)+ \Delta \phi$ is the relative phase among the three waves with the linear phase from dispersion $\phi_{\beta}$. Again, the numerical calculation show good agreement with the experimental data.

% ================================================================================================
% ================================================================================================

In summary, we experimentally demonstrated PSA based on pump-degenerate FWM in a record short 196~$\mu$m photonic chip, in spite the restriction of the maximum gain by TPA. We emphasize these are only first results and several aspects can be further optimised including: shorter pulse duration to reduce free carriers and improve the NL phase shift, improving the linear transmission, and operation closer to the zero dispersion point. We achieved an extinction ratio of 11.5~dB for both the signal and idler beams despite TPA and associated free-carriers without having to resort to P-I-N junctions, demonstrating useful levels of on-chip PSA for all-optical classic and quantum processing. 

This work was supported by the Australian Research Councils, Laureate Fellowship (FL120100029),Center of Excellence CUDOS (CE110001018) and Discovery Early Career Researcher (DE120101329, DE120102069) schemes. The authors thank Andrea Blanco-Redondo for discussion in simulation. 

% RUN EACH BibTEX SEPARATELY -- Optics Letters requires both for peer review
%\bibliographystyle{ol}    %short form style
%\bibliography{References}

\begin{thebibliography}{10}
\newcommand{\enquote}[1]{``#1''}

\bibitem{PSAinfiber2004}
C.~McKinstrie and S.~Radic, \enquote{Phase-sensitive amplification in a fiber,} Opt. Express \textbf{12}, 4973 (2004).

\bibitem{slavik2010all}
R.~Slav{\'\i}k, F.~Parmigiani, J.~Kakande, C.~Lundstr{\"o}m, M.~Sj{\"o}din,
  P.~A. Andrekson, R.~Weerasuriya, S.~Sygletos, A.~D. Ellis,
  L.~Gr{\"u}ner-Nielsen, D.~Jakobsen, S.~Herstr{\o}m, 
  R.~Phelan, J.~O'Gorman, A.~Bogris, D. Syvridis, S.~Dasgupta, P.~Petropoulos, and D.~J.Richardson, \enquote{All-optical phase and amplitude
  regenerator for next-generation telecommunications systems,} Nat. Photon.
  \textbf{4}, 690 (2010).

\bibitem{umeki2013line}
T.~Umeki, M.~Asobe, and H.~Takenouchi, \enquote{In-line phase sensitive
  amplifier based on ppln waveguides,} Opt. express \textbf{21}, 12077 (2013).

\bibitem{tong2011towards}
Z.~Tong, C.~Lundstr{\"o}m, P.~Andrekson, C.~McKinstrie, M.~Karlsson,
  D.~Blessing, E.~Tipsuwannakul, B.~Puttnam, H.~Toda, and
  L.~Gr{\"u}ner-Nielsen, \enquote{Towards ultrasensitive optical links enabled
  by low-noise phase-sensitive amplifiers,} Nat. Photon. \textbf{5},
  430 (2011).

\bibitem{levandovsky1999}
D.~Levandovsky, M.~Vasilyev, and P.~Kumar, \enquote{Amplitude squeezing of
  light by means of a phase-sensitive fiber parametric amplifier,} Opt. Lett. \textbf{24}, 984 (1999).

\bibitem{caves1982quantum}
C.~M. Caves, \enquote{Quantum limits on noise in linear amplifiers,} Phys. Rev.D \textbf{26}, 1817 (1982).

\bibitem{mcguinness2010}
H.~J. McGuinness, M.~G. Raymer, C.~J. McKinstrie, and S.~Radic, enquote{Quantum frequency translation of single-photon states in a photonic
  crystal fiber,} Phys. Rev. Lett. \textbf{105}, 093604 (2010).

\bibitem{clark2013QST}
A.~S. Clark, S.~Shahnia, M.~J. Collins, C.~Xiong, and B.~J. Eggleton, \enquote{High-efficiency frequency conversion in the single-photon regime,} Opt. Lett. \textbf{38}, 947 (2013).

\bibitem{lee2009phase}
K.-J. Lee, F.~Parmigiani, S.~Liu, J.~Kakande, P.~Petropoulos, K.~Gallo, and
  D.~J. Richardson, \enquote{Phase sensitive amplification based on quadratic
  cascading in a periodically poled lithium niobate waveguide,} Opt. Express
  \textbf{17}, 20393 (2009).

\bibitem{puttnam2011phase}
B.~J. Puttnam, D.~Mazroa, S.~Shinada, and N.~Wada, \enquote{{Phase-squeezing
  properties of non-degenerate PSAs using PPLN waveguides},} Opt. Express
  \textbf{19}, B131 (2011).

\bibitem{kang2013experimental}
N.~Kang, A.~Fadil, M.~Pu, H.~Ji, H.~Hu, E.~Palushani, D.~Vukovic, J.~Seoane,
  H.~Ou, K.~Rottwitt, and C.~Peucheret, \enquote{Experimental demonstration of
  phase sensitive parametric processes in a nano-engineered silicon waveguide,}
  in \enquote{CLEO: Science and Innovations,}  CLEO: Science and Innovations, OSA, (2013).

\bibitem{neo2013phase}
R.~Neo, J.~Schr{\"o}der, Y.~Paquot, D.-Y. Choi, S.~Madden, B.~Luther-Davies,
  and B.~J. Eggleton, \enquote{Phase-sensitive amplification of light in a
  $\chi ^{(3)}$ photonic chip using a dispersion engineered chalcogenide ridge
  waveguide,} Opt. Express \textbf{21}, 7926 (2013).

\bibitem{leuthold2010}
J.~Leuthold, C.~Koos, and W.~Freude, \enquote{Nonlinear silicon photonics,} Nat. Photon. \textbf{4}, 535 (2010).

\bibitem{politi2009integrated}
A.~Politi, J.~Matthews, M.~G. Thompson, and J.~L. O'Brien, \enquote{Integrated
  quantum photonics,} IEEE J. Sel. Top. in Quantum Electron. 
  \textbf{15}, 1673 (2009).

\bibitem{eggleton2011chalcogenide}
B.~J. Eggleton, B.~Luther-Davies, and K.~Richardson, \enquote{Chalcogenide
  photonics,}  Nat. Photon. \textbf{5}, 141 (2011).

\bibitem{yin2007impact}
L.~Yin and G.~P. Agrawal, \enquote{Impact of two-photon absorption on
  self-phase modulation in silicon waveguides,} Opt. Lett. \textbf{32}, 2031(2007).

\bibitem{foster2006broad}
M.~A. Foster, A.~C. Turner, J.~E. Sharping, B.~S. Schmidt, M.~Lipson, and A.~L.
  Gaeta, \enquote{Broad-band optical parametric gain on a silicon photonic
  chip,} Nature \textbf{441}, 960 (2006).

\bibitem{li2008systematic}
J.~Li, T.~P. White, L.~O’Faolain, A.~Gomez-Iglesias, and T.~F. Krauss, \enquote{Systematic design of flat band slow light in photonic
  crystal waveguides,} Opt. Express \textbf{16}, 6227 (2008).

\bibitem{monat2010slow}
C.~Monat, B.~Corcoran, D.~Pudo, M.~Ebnali-Heidari, C.~Grillet, M.~D. Pelusi,
  D.~J. Moss, B.~J. Eggleton, T.~P. White, L.~O'Faolain, T.~F. Krauss, \enquote{Slow light enhanced nonlinear optics in silicon photonic crystal
  waveguides,} 
IEEE J. Sel. Top. in Quantum Electron.
  \textbf{16}, 344 (2010).

\bibitem{li2011FWM}
J.~Li, L.~O'Faolain, I.~H. Rey, and T.~F. Krauss, \enquote{Four-wave mixing in
  photonic crystal waveguides: slow light enhancement and limitations,}  Opt. Express  (2011).

\bibitem{monat2009}
C.~Monat, B.~Corcoran, M.~Ebnali-Heidari, C.~Grillet, B.~J. Eggleton, T.~P.
  White, L.~O’Faolain, and T.~F. Krauss, \enquote{Slow light enhancement of
  nonlinear effects in silicon engineered photonic crystal waveguides,} Opt.
  Express \textbf{17}, 2944 (2009).

\bibitem{husko2013soliton}
C.~A. Husko, S.~Combri{\'e}, P.~Colman, J.~Zheng, A.~De~Rossi, and C.~W. Wong,  \enquote{Soliton dynamics in the multiphoton plasma regime,}
  Scientific Reports \textbf{3} (2013).

\end{thebibliography}

\bibliographystyle{osajnl} %long form style

\newpage 
\bibliographystyle{osajnl} %long form style

\end{document}